\documentstyle[manuscript,eqsecnum,aps]{revtex}                                    
\begin{document}

\draft

\title{An Effective Action for Finite Temperature Lattice
Gauge Theories with Dynamical Fermions}

\author{Peter N. Meisinger and Michael C. Ogilvie\\}
\address{Department of Physics, Washington University, St. Louis, MO 63130}

\date{\today}

\maketitle

\begin{abstract}
Dynamical fermions induce via the fermion determinant a gauge-invariant
effective action. In principle, this effective action can be added to the usual
gauge action in simulations, reproducing the effects of closed fermion loops.
Using lattice perturbation theory at finite temperature, we compute for
staggered fermions the one-loop fermionic corrections to the spatial and
temporal plaquette couplings as well as the leading $Z_N$ symmetry breaking
coupling.  A. Hasenfratz and T. DeGrand have shown that $\beta_c$ for dynamical
staggered fermions can be accurately estimated by the formula
$\beta_c = \beta^{\rm pure}_c - \Delta\beta_F$ where $\Delta\beta_F$ is the
shift induced by the fermions at zero temperature.  Numerical and analytical
results indicate that the finite temperature corrections to the zero-temperature
calculation of A. Hasenfratz and T. DeGrand are small for small values of
$\kappa = {1\over 2m_F}$, but become significant for intermediate values of
$\kappa$. The effect of these finite temperature corrections is to ruin the
agreement of the Hasenfratz-DeGrand calculation with Monte Carlo data. We
argue, however, that the finite temperature corrections are suppressed
nonperturbatively at low temperatures, resolving this apparent disagreement.
The $Z_N$ symmetry breaking coupling is small; we argue that it changes the
order of the transition while having little effect on the critical value of
$\beta$.
\end{abstract}

\pacs{PACS number(s): 12.38.G, 12.38.Mh, 11.10.Wx}

\section{INTRODUCTION}
\label{s1}
It has been known for some time that the effects of heavy dynamical fermions can
be included in Monte Carlo simulations by a hopping parameter expansion of the
fermion determinant. This is reminiscent of the Euler-Heisenberg Lagrangian of
perturbative QED, in which the effects of electron loops are included in a
gauge-invariant effective Lagrangian. Recently Hasenfratz and DeGrand
\cite{HaDe94,Lat93} have performed a zero-temperature calculation of the shift
in the lattice gauge coupling constant $\beta$, defined as $\beta=2N_c/g^2$,
induced by staggered dynamical fermions and applied the result to the finite
temperature phase transition in QCD. Their result for the shift in the critical
coupling, in the form $\beta_c = \beta^{\rm pure}_c - \Delta\beta_F$, was found
to hold rather well down to small values of the fermion mass. It is convenient
to work with a hopping parameter $\kappa$ defined by $\kappa={1 \over{2 m_F }}$,
where $m_F$ is the fermion mass. As shown in Fig.\ \ref{f1}, the
Hasenfratz-DeGrand results for $N_t = 4$ are in excellent agreement out to at
least $\kappa =2 $, and in reasonable agreement at $\kappa = 5$; this is
particularly surprising since $\Delta\beta_F$ is calculated using lattice
perturbation theory at zero temperature. In order to understand the effects of
finite temperature, we have calculated the one-loop fermionic corrections to the
spatial and temporal plaquette couplings, as well as the leading $Z_N$ symmetry
breaking coupling.

\section{RENORMALIZATION OF $\beta$ AT FINITE TEMPERATURE}
\label{s2}

\subsection{Perturbation Theory for $\Delta \beta$}
\label{ss2.1}
The $O(A^2)$ term in the
gauge field lattice action including the one-loop finite temperature fermionic
correction is given by
\begin{equation}
S_{\rm eff} = - g^2 \sum_{p_0} {1\over {N_t}} \int_{-\pi}^{~\pi} {d^3\vec{p}
\over {(2\pi)^3}} {\rm Tr}_c \Big\{ \tilde{A}_{\mu}(p) \tilde{A}_{\nu}(p) 
{\beta \over{2N_c}} \Bigl[ D_{\mu\nu}^{(0)}(p) + D_{\mu\nu}^{(1)}(p) 
- D_{\mu\nu}^{(2)} \Bigr] \Big\}
\label{e2.1.1}
\end{equation}
where
\begin{equation}
D_{\mu\nu}^{(0)}(p) = 4 \Bigl[\delta_{\mu\nu} \sum_{\alpha}
\sin^2\Bigl({p_{\alpha} \over 2}\Bigr) - \sin\Bigl({p_{\mu} \over 2}\Bigr)
\sin\Bigl({p_{\nu} \over 2}\Bigr)\Bigr],
\label{e2.1.2}
\end{equation}
\begin{equation}
D_{\mu\nu}^{(1)}(p) = {1 \over 2} \sum_{k_0} {1 \over {N_t}} \int_{-\pi}^{~\pi}
{d^3\vec(k) \over {(2\pi)^3}} {\rm Tr}_d \Bigl[ R \Bigl( k_{\mu}+
{p_{\mu} \over 2} \Bigr) S^{-1}(k)
R \Bigl( k_{\nu}+{p_{\mu} \over 2} \Bigr) S^{-1}(k+p) \Bigr],
\label{e2.1.3}
\end{equation}
and
\begin{equation}
D_{\mu\nu}^{(2)} = {1 \over 2} \delta_{\mu\nu} \sum_{k_0} {1 \over {N_t}}
\int_{-\pi}^{~\pi} {d^3\vec(k) \over {(2\pi)^3}} {\rm Tr}_d
\Bigl[ Q(k_{\mu}) S^{-1}(k) \Bigr]
\label{e2.1.4}
\end{equation}
with the vertex functions given by
\begin{equation}
R(k_{\mu}) = i\gamma_{\mu}\cos(k_{\mu})
\label{e2.1.5}
\end{equation}
and
\begin{equation}
Q(k_{\mu}) = -i\gamma_{\mu}\sin(k_{\mu})
\label{e2.1.6}
\end{equation}
with no sum over $\mu$. The inverse fermion propagator is
\begin{equation}
S(k) = {1 \over {2\kappa}} + i\gamma_{\mu}\sin(k_{\mu}).
\label{e2.1.7}
\end{equation}
This formula is a straightforward consequence of the lattice Feynman rules,
which are given in Fig.\ \ref{f2}. The diagrams contributing to the fermionic
renormalization of $\Delta \beta$ are shown in Fig.\ \ref{f3}. The first
diagram, corresponding to $D^{(0)}$, is the free lattice gluon propagator. The
second diagram, corresponding to $D^{(1)}$, involves R and S only and survives
in the continuum limit; note that R is the lattice analog of the continuum gluon
vertex. The third diagram, corresponding to $D^{(2)}$, is a lattice tadpole
diagram, and involves the vertex function Q, which is a feature only of the
lattice theory. At zero temperature, this tadpole contribution vanishes after
integration by parts \cite{CeMa81}. Finite temperature enters into the
calculation only through the replacement of the integration over the $k_0$
variable appropriate for zero temperature by the sum over Matsubara frequencies
\begin{equation}
k_0 = \frac{2 \pi n }{T}
\label{e2.1.8}
\end{equation}
where n is integer-valued and T is the temperature.

\subsection{Ward Identity at Finite Temperature}
\label{ss2.2}
At finite temperature, the $D^{(2)}$ term in Eq. (\ref{e2.1.1}) is necessary in
order to show that the lattice form of the Ward identity
\begin{equation}
\sin \Bigl( {p_{\nu} \over 2} \Bigr) \Bigl[ D_{\mu\nu}^{(2)}(p) -
D_{\mu\nu}^{(1)} \Bigr] = 0
\label{e2.2.1}
\end{equation}
still holds at finite temperature, even though the four dimensional hypercubic
symmetry is broken. To show this, we first note the two identities
\begin{equation}
S(k+p) - S(k)
= 2 \sin \Bigl( {p_{\mu} \over 2} \Bigr) R \Bigl( k_\mu+{p_{\mu} \over 2}
\Bigr)
\label{e2.2.2}
\end{equation}
and
\begin{equation}
R \Bigl( k_\mu + {p_\mu \over 2} \Bigr) - R \Bigl( k_\mu - {p_\mu \over 2}
\Bigr) = 2 \sin \Bigl( {p_{\mu} \over 2} \Bigr) Q(k_\mu).
\label{e2.2.3}
\end{equation}
Use of the first identity gives
\begin{eqnarray}
\sin \Bigl( {p_{\nu} \over 2} \Bigr) \Bigl[ D_{\mu\nu}^{(2)}(p) -
D_{\mu\nu}^{(1)} \Bigr] =
&&\frac{1}{4} \sum_k \frac{1}{N_t V } Tr_d
\Bigl\{ 2 \sin \Bigl( \frac{p_\mu}{2} \Bigr) Q( k_\mu )
S^{-1}(k) \nonumber\\
&&+ \Bigl[ R \Bigl( k_\mu+{p_{\mu} \over 2} \Bigr) S^{-1}(k_{\mu} + p_{\mu}) -
R \Bigl( k_\mu+{p_{\mu} \over 2} \Bigr) S^{-1}(k_\mu)
\Bigr] \Bigr\},
\label{e2.2.4}
\end{eqnarray}
and after a simple shift of variables, use of the second identity yields the
desired cancellation.

\subsection{Finite Temperature Decomposition of the Propagators}
\label{ss2.3}
At finite temperature there are two independent symmetric tensors of order $p^2$
which are four-dimensionally transverse\cite{KaKa85}. In what follows, all
expressions will be in the thermal rest frame.  The first of the corresponding
lattice tensors is specified by
\begin{equation}
P_{\mu 0}^{(3)} (p) = P_{0 \mu}^{(3)} (p) = 0
\label{e2.3.1}
\end{equation}
\begin{equation}
P_{ i j }^{(3)} (p) = \delta_{ i j }
- {{\tilde{p}_i \tilde{p}_j}  \over { \tilde{p}_{\it s}^2}}
\label{e2.3.2}
\end{equation}
and the second is
\begin{equation}
P_{\mu \nu}^{(4)} (p) - P_{\mu \nu}^{(3)} (p)
\label{e2.3.3}
\end{equation}
where
\begin{equation}
P_{\mu \nu}^{(4)} (p) = \delta_{\mu \nu}
- {{\tilde{p}_\mu \tilde{p}_\nu } \over { \tilde{p}^2}}
\label{e2.3.4}
\end{equation}
The lattice quantities $\tilde{p}$ are defined by
\begin{equation}
\tilde{p}_\mu = 2 \sin \Bigl( {p_{\mu} \over 2} \Bigr)
\label{e2.3.5}
\end{equation}
\begin{equation}
\tilde{p}^2 = \sum_\mu \tilde{p}_\mu^2
\label{e2.3.6}
\end{equation}
\begin{equation}
\tilde{p}_s^2 = \sum_i \tilde{p}_i^2
\label{e2.3.7}
\end{equation}

The existence of these two independent tensors leads to separate
renormalizations of the spatial and temporal gauge couplings at finite
temperature. The first tensor, $P_{\mu \nu}^{(3)} (p)$, is associated with the
magnetic, or spatial, part of the action, while the second tensor,
$P_{\mu \nu}^{(4)} (p) - P_{\mu \nu}^{(3)} (p)$, is associated with the
electric, or temporal, part of the action.

We find that
\begin{equation}
\Delta\beta_s = - N_c \sum_{k_0} {1\over {N_t}} \int_{-\pi}^{~\pi} {d^3\vec{k}
\over {(2\pi)^3}} \Phi(k;1,2)
\label{e2.3.8}
\end{equation}
and
\begin{equation}
\Delta\beta_t = - N_c \sum_{k_0} {1\over {N_t}} \int_{-\pi}^{~\pi} {d^3\vec{k}
\over {(2\pi)^3}} \Phi(k;0,1)
\label{e2.3.9}
\end{equation}
where
\begin{eqnarray}
\Phi(k;\mu,\nu) = &&32B^{-2}(k)\cos^2(k_{\mu})\cos^2(k_{\nu}) \nonumber\\ 
&&- 4096B^{-4}(k) \sin^2(k_{\mu})\cos^2(k_{\mu}) \sin^2(k_{\nu})\cos^2(k_{\nu})
\label{e2.3.10}
\end{eqnarray}
and
\begin{equation}
B(k) = {1 \over {\kappa^2}} + 4 \sum_{\alpha} \sin^2(k_{\alpha}).
\label{e2.3.11}
\end{equation}
As the temperature is taken to zero, the two expressions smoothly approach
each other to give the zero-temperature result.

\subsection{Numerical Results for $\Delta \beta$}
\label{ss2.4}
The integrals (\ref{e2.3.8}) and (\ref{e2.3.9}) were evaluated numerically by
calculating mode sums for large values of $N_s$ and various values of $N_t$.
Figure \ref{f4} compares $\Delta\beta$ per fermion (spatial and temporal) vs.
$\kappa$ for $N_t = 4$ and $N_t = 6$ with the zero temperature result of
Hasenfratz and DeGrand \cite{HaDe94}. The finite temperature values approach
the zero temperature result from below as a consequence of the antiperiodic
boundary conditions; as $N_t$ increases, the sum includes more terms in the
region near $p = 0$, which dominates for small $m_F$. As might be expected, the
temporal shift in $\beta$ is more sensitive to the effect of finite temperature
than the spatial shift.  For small values of $\kappa$, corresponding to large
values of the fermion mass, the effects of finite temperature are small. This is
easily understood, since $\kappa$ is much smaller than $N_t$. However, for
intermediate values of $\kappa$, finite temperature corrections ruin the
excellent agreement between the zero-temperature calculation and the Monte Carlo
results discussed in Sec.\ \ref{s1}.

\subsection{Image Expansion}
\label{ss2.5}
The connection between the the zero and finite temperature result can be
understood more physically by transforming the sum over Matsubara frequencies
into a sum over images using the Poisson summation formula for antiperiodic
boundary conditions
\begin{equation}
\sum_{p_0} {1 \over {N_t}} F(p_0) = \sum_n (-1)^n \int_{-\pi}^{~\pi}
{dp_0 \over {2\pi}} F(p_0) e^{inN_tp_0}
\label{e2.5.1}
\end{equation}
so that, for example, the shift in the spatial coupling is given by
\begin{equation}
\Delta\beta_s = - N_c \int_{-\pi}^{~\pi} {d^4k \over {(2\pi)^4}} \Phi(k;1,2)
- 2 N_c \sum_{n=1}^{\infty} (-1)^n \int_{-\pi}^{~\pi} {d^4k
\over {(2\pi)^4}} \Phi(k;1,2) \cos(nN_tk_0)
\label{e2.5.2}
\end{equation}
with a similar result for $\Delta\beta_t$. This form has a simple physical
interpretation: the first integral is the zero-temperature shift, and the
integer $n$ in the second term labels the net number of times the fermion wraps
around the lattice in the temporal direction. The finite temperature corrections
result from the $O(A^2)$ expansion of image diagrams such as those depicted in
Fig.\ \ref{f5}. Numerically, the dominant corrections to the zero-temperature
result come from the first few values of $n$, with the $n = 1$ and $n = 2$ terms
accounting for more than $90\%$ of the finite temperature correction for
$\kappa \leq 2.0$.

Although not apparent in our perturbative calculations, in order to maintain
gauge invariance, the vertical segments of the image diagrams must be
accompanied by powers of Polyakov loops. It is an observed feature of
simulations with dynamical fermions that the $Z_N$ symmetry is approximately
maintained in the low-temperature phase. This suggests that the image
contibutions may be negligible below $\beta_c$. Thus, the zero temperature
corrections to $\beta$ are suppressed nonperturbatively in the confined regime.
In particular, just below $\beta_c$ the zero-temperature result will hold for
$\Delta\beta$. Figure \ref{f6} illustrates an idealized behaviour for
$\Delta\beta_{\rm fermion}$ as a function of $\beta_{\rm pure}$.

\section{$Z_N$ SYMMETRY BREAKING IN THE EFFECTIVE ACTION}
\label{s3}
There is another set of terms induced by the fermion determinant only at finite
temperature. As is well known, the $Z_N$ symmetry of the pure gauge theory is
explicitly broken by dynamical fermions. To lowest order in the hopping
parameter expansion, a path of $N_t$ hops around the lattice in the temporal
direction produces a effective coupling to the Polyakov loop, explicitly
breaking the $Z_N$ symmetry. Evaluating the fermion determinant in
a constant $A_0$ background field and applying the 
Poisson summation formula [Eq.\ (\ref{e2.5.1})] once again, we find an
additional contribution to the effective action:
\begin{equation}
S_{\rm eff} = \sum_{\vec{x}} \sum_{n=1}^{\infty} (-1)^{n+1} h_n(\kappa)
Re \Bigl[ {\rm Tr} P^n(\vec{x}) \Bigr]
\label{e3.1}
\end{equation}
where $P(\vec{x})$ denotes a Polyakov loop and the couplings $h_n$ are given by
\begin{equation}
h_n(\kappa) = -4 N_t \int_{-\pi}^{~\pi} {d^4q \over {(2\pi)^4}} {\rm ln} \Biggl[
{1 \over {4\kappa^2}} + \sum_{\mu} \sin^2(q_{\mu}) \Biggr] \cos(nN_tq_0)
\label{e3.2}
\end{equation}
This result can also be obtained by using a contour integral to evaluate the
sum over Matsubara frequencies. The leading term in this effective action has
been discussed for the case $N_t = 2$ \cite{HaKaSt83}.

\subsection{Numerical Results for h}
\label{ss3.1}
The maximum values of the $h_n$ are obtained when $m_F = 0$. For $N_t = 4$ we
find $h_1^{\rm max} = 0.107$, $h_2^{\rm max} = 0.00445$, and increasingly
smaller magnitudes for higher order couplings. Because $h_1$ favors $Z_N$
breaking, it acts to lower the critical value of $\beta$. Unlike the
$D^{(1)}(p)$ term considered in Sec.\ \ref{s2}, this effect cannot be directly
included as a finite shift in $\beta$. The most direct way to determine the
shift in $\beta$ due to $h_1$ term is to perform a Monte Carlo simulation of the
pure gauge theory with an $h_1$ term added to the action. The numerical
simulation results presented in this section were obtained from runs of 40,000
sweeps (after thermalization) on a $10^3\times4$ lattice using a variety of work
stations.

We have observed that this additional source of $\Delta\beta$ is small compared
to the renormalization of the plaquette couplings discussed in the preceeding
section, regardless of $m_F$. For example, even $h_1 = 0.1$ at $N_t = 4$ leads
to a shift in $\beta$ per fermion of $0.00325$. This value of $h_1$ corresponds
to $m_F = 0.17$ which yields a zero-temperature predicted shift in $\beta$ per
fermion of $0.104$. Although the effect of the $h_1$ term in the effective
action on the critical value of $\beta$ is quite small, $h_1$ has a profound
effect on the character of the transition. Figure~\ref{f7} shows the frequency
distribution for the spatial expectation value of the Polyakov loop as a
function of $h_1$ at the appropriate $\beta_c (h_1)$ on a $10^3 \times 4$
lattice. As $h_1$ increases, the peaks associated with the two phases move
closer together until they merge at what is presumably a second-order critical
end-point. A first-order phase transition exists for values of $h_1$ smaller
than approximately $0.08$.

The Monte Carlo results for the phase diagram of a pure gauge theory with an
$h_1$ coupling can be mapped back to the phase diagram of QCD with dynamical
quarks by including both the $h_1$ term and the shift $\Delta \beta$. Including
zero-temperature plaquette coupling renormalization, the endpoint of this
first-order phase transition line maps to the point $(0.394, 4.68)$ in the
$(m_F, \beta)$-plane for the case of sixteen degenerate staggered fermion
species. This is roughly consistent with the endpoint of the first-order phase
transition observed in simulations with dynamical staggered fermions
\cite{OhKi91}. However, a definitive comparison will likely require significant
computational resources.

\section{CONCLUSIONS}

Figure \ref{f8}, a graph of $\beta_c$ versus $\kappa$, summarizes the results
we have obtained. As shown by Hasenfratz and DeGrand, the zero-temperature shift
in the coupling constant due to dynamical fermions nicely accounts for the shift
in $\beta_c$. We have found that finite temperature corrections to the gauge
coupling renormalization lift the degeneracy of the spatial and temporal
couplings, and the results are significantly different from the zero-temperature
results. As can be seen from the figure, they are in conflict with the Monte
Carlo data. At finite temperature, dynamical fermions couple to Polyakov loops
via loops circling the lattice in the timelike direction. This also acts to
shift the value of $\beta_c$. This shift is small, however, and does not restore
the success of the zero-temperature calculation.

As we have shown, the image expansion makes it plausible that the success of the
zero-temperature calculation in determining the critical value of $\beta$ is due
to a nonperturbative suppression of finite temperature effects in the low
temperature regime. Specifically, the small expectation value of the Polyakov
loops at low temperatures indicates a suppression of those quark paths which
account for finite temperature corrections. This leads to the behavior of the
effective gauge coupling constants shown in Fig.\ \ref{f5} above. If this
picture is correct, a gauge theory with dynamical fermions of intermediate mass
is well modeled by a pure gauge theory in which the spatial and temporal
couplings are slightly different. Along the critical line, both couplings jump
discontinuously. This could be studied in more detail by adjusting spatial and
temporal couplings in a pure gauge theory simulation so that plaquette
expectation values matched those observed in dynamical simulations. This might
also afford an opportunity to consider the effects of finite spatial sizes. Such
effects are easily included in Eq.\ \ref{e2.5.2} for the coupling shift by
choosing the spatial mode sums appropriately.

While the coupling to Polyakov loops induced by dynamic fermion loops seems to
play little role in determining $\beta_c$, this coupling does influence the
order of the transition; in our $N_t = 4$ simulations of a pure gauge theory
with an additional coupling $h_1$ to Polyakov loops, a sufficiently large value
of $h_1$ causes the first order line to terminate, while shifting $\beta_c$ very
little. Thus, the effective coupling $h_1$ appears to be the most important
factor in determining the end point of the first-order deconfining phase
transition line.

In addition to the transition temperature, other quantities can be estimated
using these perturbative techniques. For example, the chiral order parameter
$\langle \bar{\psi} \psi \rangle$ can be estimated as a constant term plus a
term proportional to the expectation value of a plaquette plus a term
proportional to the expectation value of the Polyakov loop. However, the use of
a perturbative evaluation of the fermion determinant obviously fails to include
the effects of chiral symmetry breaking, which is the dominant factor in
determining $\langle \bar{\psi} \psi \rangle$ for light quarks. Presumably this
accounts for the failure of the effective theory for light quark masses.

\section{ACKNOWLEDGEMENTS}
We would like to thank the U.S. Department of Energy for financial support under
Contract No. DE-AC02-76-CH00016. One of us (PNM) would also like to thank the
U.S. Department of Education for additional financial support in the form of a
GANN Predoctoral Fellowship.

\begin{figure}
\caption{Predicted values (circles) of $\Delta\beta$ per fermion at zero
temperature versus $\kappa$ compared with finite temperature simulation results
(squares).}
\label{f1}
\end{figure}

\begin{figure}
\caption{Lattice Feynman rules to ${\cal O}(A^2)$. The Q vertex is not present
in the continuum limit.}
\label{f2}
\end{figure}

\begin{figure}
\caption{One-loop diagrams contributing to the fermionic renormalization of
$\Delta\beta$.}
\label{f3}
\end{figure}

\begin{figure}
\caption{Spatial and temporal values of $\Delta\beta$ per fermion versus
$\kappa$ for $N_t = 4, 6, and \infty$.}
\label{f4}
\end{figure}

\begin{figure}
\caption{Typical diagrams from the expansion of $\Delta\beta$ in images.}
\label{f5}
\end{figure}

\begin{figure}
\caption{Idealized behaviour of $\Delta\beta_{\rm fermion}$ as a function of
$\beta_{\rm pure}$.}
\label{f6}
\end{figure}

\begin{figure}
\caption{Frequency distribution for the expectation value of the Polyakov loop
on a $10^3 \times 4$ lattice at $\beta_c(h_1)$ for $h_1$~=~0.025, 0.05, 0.075,
and 0.1.}
\label{f7}
\end{figure}

\begin{figure}
\caption{$\Delta\beta$ per fermion versus $\kappa$ computed at zero temperature,
at finite temperature, and at finite temperature with a correction for Polyakov
loop effects compared with finite temperature simulation results.}
\label{f8}
\end{figure}


\begin{references}

\bibitem{HaDe94} A.~Hasenfratz and T.~DeGrand, Phys. Rev. D {\bf 49}, 466
(1994).
\bibitem{Lat93}  A.~Hasenfratz and T.~DeGrand, Nucl. Phys. B (Proc. Suppl.)
{\bf 34}, 317 (1994).
\bibitem{CeMa81} W.~Celmaster and D.~Maloof, Phys. Rev. D {\bf 24}, 2730
(1981).
\bibitem{KaKa85} K. Kajantie and J. Kapusta, Ann. Phys. {\bf 160}, 477 (1985).
\bibitem{HaKaSt83} P.~Hasenfratz, F.~Karsch, and I.~O.~Stamatescu, Phys. Lett.
{\bf 133B}, 221 (1983).
\bibitem{OhKi91} S.~Ohta and S.~Kim, Phys. Rev. D {\bf 44}, 504 (1991).

\end{references}
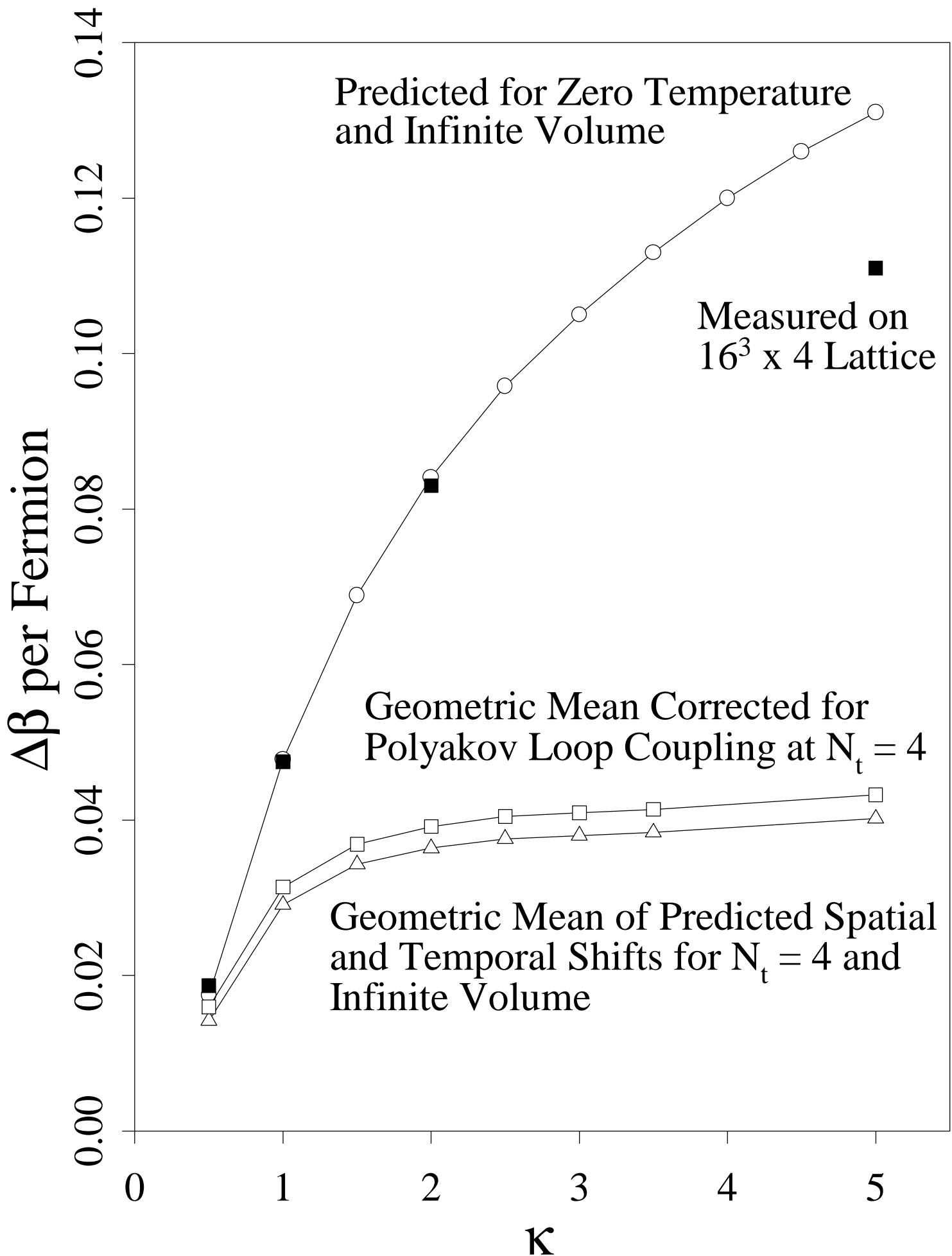
\end{document}